\documentclass[showpacs,amsmath,amssymb,10pt,twocolumn,prl]{revtex4}

\usepackage{graphicx}% Include figure files
\usepackage{dcolumn}% Align table columns on decimal point
\usepackage{bm}% bold math
\usepackage{epsfig}

\newcommand{\gd}{\delta}
\newcommand{\gre}{\varepsilon}

\newcommand{\gm}{\mu}
\newcommand{\gmh}{\hat \mu}

\newcommand{\gs}{\sigma}

\newcommand{\go}{\omega}

\newcommand{\gD}{\Delta}

\newcommand{\gS}{\Sigma}

\newcommand{\gO}{\Omega}

\newcommand{\oN}{{\mathbb N}}

\newcommand{\oR}{{\mathbb R}}

\newcommand{\barx}{{\bar x}}
\newcommand{\bargs}{{{\bar \gs}^2}}

\newtheorem{theorem}{Theorem}
\newtheorem{lemma}[theorem]{Lemma}

\newtheorem{prop}{Proposition}[section]
\newtheorem{claim}[theorem]{Claim}

\newtheorem{definition}[theorem]{Definition}
\newtheorem{question}{Question}[section]
\newtheorem{coro}{Corollary}[section]

\newcommand{\beq}{\begin{equation}}
\newcommand{\eeq}{\end{equation}}
\newcommand{\bea}{\begin{array}}
\newcommand{\ena}{\end{array}}
\newcommand{\bds}{\begin {itemize}}
\newcommand{\eds}{\end {itemize}}
\newcommand{\bdf}{\begin{definition}}
\newcommand{\blm}{\begin{lemma}}
\newcommand{\edf}{\end{definition}}
\newcommand{\elm}{\end{lemma}}
\newcommand{\bthm}{\begin{theorem}}
\newcommand{\ethm}{\end{theorem}}
\newcommand{\bprp}{\begin{prop}}
\newcommand{\eprp}{\end{prop}}
\newcommand{\bcl}{\begin{claim}}
\newcommand{\ecl}{\end{claim}}
\newcommand{\bcr}{\begin{coro}}
\newcommand{\ecr}{\end{coro}}
\newcommand{\bquest}{\begin{question}}
\newcommand{\equest}{\end{question}}

\newcommand{\rarrow}{{\rightarrow}}

\newcommand{\restrict}{{\upharpoonright}}
\newcommand{\nin}{{\not \in}}

\newcommand{\round}{\hbox{round}}

\newcommand{\bbN}{{\mathbb N}}

\begin{document}
%

%\preprint{APS/MPML-001}
\title{An experimental uncertainty implied by failure of the physical
 Church-Turing thesis}
\author{Amir Leshem}
\affiliation{School of Engineering, Bar-Ilan university,\\ 52900,
Ramat-Gan, Israel}
\homepage{http://www.eng.biu.ac.il/~leshema} 
\date{\today}% It is always \today, today,
\begin{abstract}
In this paper we prove that given a black box assumed to generate bits
of a given non-recursive real $\Omega$  there is no computable decision
procedure generating sequences of decisions such that if the output is
indeed $\Omega$ the process eventually accepts the hypothesis while if the
output is different than $\Omega$ than the procedure will eventually
reject the hypothesis from a certain point on. Our decision concept 
does not require full certainty regarding the correctness of the decision at any point, thus better represents the validation process of physical theories.
The theorem has strong implications on the falsifiability of physical theories entailing the failure of the physical Church Turing thesis. Finally we show that our decision process enables to decide whether the mean of an i.i.d. sequence of reals belongs to a specific $\Delta_2$ set of integers. This significantly strengthens the effective version of the Cover-Koplowitz theorem, beyond computable sequences of reals.
\end{abstract}
%\keywords{Church-Turing thesis, Hypothesis testing, experimental uncertainty}
\maketitle
\section{Introduction}
Church-Turing thesis states that every function $f:\bbN \rarrow \bbN$
that is algorithmically computable is recursive or equivalently
computable by a universal Turing machine. This is a mathematical
statement about algorithms not physical computers. 
 This thesis has been supported by the equivalence of all models of 
computations to date, e.g., Turing machines, register machines
and lambda calculus \cite{enderton}. A
stronger thesis termed the physical Church-Turing thesis is that every
finitely realizable physical process can be perfectly simulated by a
universal Turing machine operating by finite means (see \cite{deutsch85}, \cite{penrose90} and the references therein). 
This thesis is much stronger. It has been argued that there might be finitely 
realizable  physical processes that can compute non-computable functions 
 e.g., by analog computation \cite{siegelmann95},  stronger versions of quantum computing \cite{nielsen97},\cite{kieu2003} or general relativity \cite{shagrir2003}. Furthermore  \cite{penrose90,penrose94} claims that 
some of the open problems in physics (e.g., quantum gravitation) and the failure to generate artificial intelligence might have underlying non-computable physical
processes.
Similarly the possibility that some physical constants are non-computable has 
been conjectured. If indeed a physical constant is
non-computable then more and more accurate measurements of the
constant (or repeated independent observations with i.i.d noise) will provide 
us more bits of its binary expansion, therefore
providing an oracle that breaks the physical Church-Turing thesis.  

In this paper we discuss the empirical validation and refutation of the hypothesis 
that a computing device generates a non-computable real number. 
Deutsch \cite{deutsch85} argues that this is 
hopeless since any finite experiment generates a finite sequence of 
measurements with finite precision so the output of any experiment is 
computable. However, following Cover \cite{cover73} we might envision a 
different strategy:  Our null hypothesis is that we are given a black box 
generating the bits of a specific non-computable real
(In order to be able to use the digits provided by the black box we need to 
know what is the assumed outcome, otherwise a random number generator will provide us  almost surely with useless bits of a non-computable real).
We now perform an infinite sequence of tests, each sampling more and 
more bits of the given black box.  This measurement procedure serves as
an oracle, and our task is to refute the null hypothesis if the oracle
is a false oracle. The only requirement we have is that if the oracle
is a false oracle (does not provide the bits of $\gO$) then from a certain 
point our procedure always rejects the null hypothesis, and if it indeed 
provides the bits of $\gO$ then from a certain point on we will always accept 
the hypothesis. This viewpoint is in line with the idea of falsifying a 
physical theory \cite{popper59}. Note that while our decisions are 
asymptotically correct we are
never certain about this fact. Since at any finite stage we might change our
decision, we never get full certainty regarding the null hypothesis, only 
growing confidence. This is 
much like the validation process of physical theories. The theory is accepted 
only if it is sufficiently simple on one hand and has not yet been refuted by 
experiment on the other hand.

Cover \cite{cover73} used this strategy to
verify that a mean of a random sequence is rational. His motivation was to 
provide tests for certain simple representations of well known physical 
constants \cite{wyler71, lenz51, good70}.  While at
every finite stage any confidence interval contains rationals and
irrationals he shows that if the number is rational or is irrational
outside a certain set of measure 0 then with probability one the
sequence of decisions is correct from a certain point on. Koplowitz 
\cite{koplowitz77} and
later Peres and Dembo \cite{peres94} proved that if we want to test a 
hypothesis regarding two sets contained in disjoint $F_{\gs}$ sets then
the correct decision is made with probability one for both $H_1$
and $H_0$ and the set of measure 0 can be assumed empty. 
However in all these papers little attention was paid to
actual computability of the decision procedures. The procedures
are computable only if the countable sequence of reals is a 
computable sequence of computable reals. 

In this paper we apply the same decision concept to testing the
non-computability of physical constants (and actually to any output of
a physical black box) providing us with a sequence of approximations
to a real number. We show that given the output of a black box which is 
assumed to provide us bits of a given non-computable real there is NO decision
procedure (deterministic or probabilistic) that makes an infinite
sequence of decisions such that any false oracle is detected from a
certain point on. The proof is general enough to apply to any physical
model of computation. It means that if the Church-Turing thesis is
true, than there is no way to experimentally refute the hypothesis that 
a finitely realizable process provides us a non-computable number. 
This type of uncertainty requires us to choose between falsifiability of
such a physical prediction (which would be the most substantial prediction of a
theory claiming that a given physical process is non-computable) 
and our belief in the given theory. Hence accepting the failure of the 
physical Church-Turing thesis implies that we need to revise the notion of
experimental refutation of physical theories as proposed by Popper.

On the positive side we provide a computable decision procedure that almost 
surely decides asymptotically whether the mean of an i.i.d random process 
belongs to a given $\gD_2$ non-computable set of natural numbers 
using the asymptotic decision concept.
Our result significantly strengthens Cover's result since 
$\gD_2$ sets cannot be effectively enumerated, and therefore Cover's proof 
fails to provide a computable decision procedure. 

\section{Physical processes, Decision procedures and experiment design}
In this section we review the concept of an experiment and a decision
procedure to falsify a given prediction of a physical theory.
We focus on the experimental evidence that the output of a black
box is a non-computable real. The formulation is sufficiently general to
include statistical decisions as well as deterministic procedures.

First we comment that basically any physical observation is subject to noise 
and finite accuracy limitations. This implies that the best we can hope from 
an experimental devices is to obtain a sequence of measurements with increasing
accuracy rather than a single infinitely accurate measurement. However we do
assume that there is no practical limitation on the observation and that by 
averaging we might reduce effects of observation noise as much as we would 
like (otherwise it would be trivially impossible to decide anything regarding 
non-computability, since any finite sequence is by itself computable). 
Even this assumption ignores some inherent limitations on measurement such as
the fact that certain parameters cannot be simultaneously measured due to 
quantum limitations, or fundamental limitations on synchronization, etc.

We now provide some definitions and notations to be used later in the
paper.
\bdf
\begin{enumerate}
\item For a finite sequence $s=\left<s(0),s(1),...,s(N-1)\right>$ we
denote the length of the sequence by $l(s)=N$.
\item For two sequences $s,t$ where $l(s) \le l(t)$ we denote
$s \le^* t$ if for all $n < l(s)$ we have $s(n)=t(n)$.
\item Let $s$ be a sequence of integers (finite or infinite). Let $n \le
l(s)$ then $s \restrict n=\left<s(0),s(1),...,s(n-1)\right>$.
\item $\go$ is the set of natural numbers and $2^{<\go}$ is the binary tree
of all binary finite sequences. 
\end{enumerate}
\edf

Let $\gO$ be a non-computable real. We shall assume without
loss of generality that our real $\gO$ is the characteristic function
of a non-computable set $W$. 
Suppose that as in figure 1 a black box S, provides us at each
experiment $n=0,1,2,...$ with a bit $S(n)$.  
We would like at each stage $n$ to decide whether the bits provided by
$S$ so far are the initial bits of our real $\gO$ or not. We do not
require that all our decisions are correct, however we require that for any
$S=<S(0),S(1),...>$, if $S \neq \gO$ then from a certain finite
stage we decide correctly that $S \neq \gO$. Similarly if $S=\gO$ then
we do not change our mind regarding this and we accept the null
hypothesis from a certain stage onwards. Note that the decision
procedure can be viewed as a computable function 
$F:2^{<\go} \rarrow \{0,1\}$ where for any $s=<s(0),...,s(N-1)>$ 
\beq
F(s)=0 \hbox{\ iff we decide that $s \not \le^* \gO$}.
\eeq

\begin{figure}
    \begin{center}
    \mbox{\psfig{figure=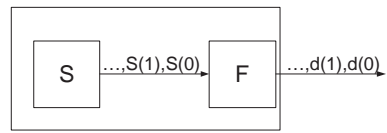,width=0.28\textwidth}}
    \end{center}
    \caption{A general decision process for refuting the null
    hypothesis $S=\gO$}
  \label{experiment}
\end{figure}

Let $S$ be a source of bits. We measure the output of $S$. 
A physical theory stating that $S$ generates the
bits of a specific non-recursive real is experimentally falsifiable if
and only if there is a computable decision procedure $F:2^{<\go}
\rarrow \{0,1\}$ such that for every $S$, $ \lim_{n \rarrow \infty} F(S \restrict n)$ exists and 
\beq
\lim_{n \rarrow \infty} F(S \restrict n)=0 \iff S \neq \gO.
\eeq
The setup is presented in figure \ref{experiment}.
We do not require that we know whether $S\neq \gO$ or not at any
finite stage since any decision might be reversed after further observation. 
However we do require that performing more and more measurements of the output 
of $S$ will eventually accept the null
hypothesis 
\beq
H_0: S=\gO
\eeq
if it is valid or reject it from a certain point on  if  $S \neq
\gO$. 
This notion of verification is consistent with the way we test
physical theories in general \cite{popper59}. We do not know at any 
finite stage whether the theory is correct but we require that the theory
is falsifiable, i.e., if the theory is wrong then there is an experiment
falsifying the theory. If the theory is false then from a certain point on our experiments will refute the theory.

\section{Main theorem}
The asymptotic verification described above 
 can be utilized in all existing physical
theories e.g., verifying that the absolute zero is not achievable. This 
is not the case with non-recursiveness. Our main theorem proves that
the non-recursiveness of the outcome of a finitely realizable physical system
implies a significant
measurement uncertainty in the sense that no experiment can be designed to
eventually refute or accept the prediction of the theory. Interestingly 
theorem \ref{delta2} provides an example of asymptotic decision process 
regarding membership in $\gD_2$ sets. Hence not all is lost regarding 
testing hypotheses about non-recursive sets. 

\bthm
\label{main_theorem}
Let $W$ be a non-recursive set (e.g., the set of G\"{o}del numbers of all 
halting Turing machines). Let $\gO=1_{W}$ be the characteristic
function of $W$. Then there is no computable decision procedure 
$F:2^{< \go} \rarrow \{0,1\}$ such that $\lim_{n \rarrow \infty} F(S \restrict n)$ exisits and  
\beq
\label{asymptotic}
\lim_{n \rarrow \infty} F(S \restrict n)=0 \iff S \neq \gO
\eeq
\ethm 
Proof: Assume towards contradiction that we are given a recursive
$F:2^{< \go} \rarrow \{0,1\}$ such that (\ref{asymptotic}) holds.
\bcl
For every $t$  define 
\[
R_t=\left\{r | t \le^*r \wedge F(r)=1 
\right\}.
\]
If $t \not \le^* \gO$ then $R_t$ is finite.
\ecl
Proof of the claim: Assume that $t \not \le^* \gO$ and 
 $R_t$ is infinite. Define a subtree $T_t$ of $2^{<\go}$ by
\[
T_t=\left\{s | t \le^*s \wedge \exists r \in R_t (s \le^* r) 
\right\}.
\]
This is the subtree spanned by all extensions of $t$ that are decided the
wrong way (since $t \not \le^* \gO$). Since $R_t$ is infinite $T_t$ is
an infinite tree with finite branching. Therefore by K\"{o}nig's lemma it
has an infinite
branch $b$. However for every node $s \le^* b$ there is an extension
$r \in b$ satisfying $F(r)=1$. Therefore 
\[
\lim_{n \rarrow \infty} F(b \restrict n) \neq 0
\] 
contradicting the definition of $F$, since if $S$ generates $b$ either we
asymptotically accept $H_0$, i.e., we decide $b=\gO$, which is false, or we 
change our decision
infinitely many times. This contradicts the definition of $F$ and the claim is 
proved.

We are now in position to finish the proof of theorem \ref{main_theorem}. Let 
$m_0$ be sufficiently large so that $F(\gO \restrict m)=1$ for all $m_0 \le
m$. 
Such an $m_0$ exists since by (\ref{asymptotic}).
By our claim we have 
\[
t \le^* \gO \iff \forall m \left(m_0 \le m \rarrow \exists s \in
2^m(t \le^* s \wedge F(s)=1) \right).
\]
Let $m_1=\max\{n,m_0\}$. Then for all $n$ 
\[
n \in W \iff 
\forall m \left( m_1 \le m \rarrow \exists s \in
2^m(F(s)=1) \wedge s(n)=1 \right).
\]
Similarly for all $n$ 
\[
n \not \in W \iff \forall m \left(m_1 \le m \rarrow 
\exists s \in 2^m(F(s)=1) \wedge s(n)=0 \right).
\]
Since all internal quantifiers are bounded $W$ is both a $\Pi_1$ and 
$\gS_1$ set so it is recursive contradicting our assumption.

\section{Deciding that a mean belongs to a non-computable set}
While in the previous section we proved that even given a deterministic
output of a black box we cannot define a computable sequence of decisions
asymptotically deciding whether the black box provides us the digit of a 
specific non-computable real, we show that certain decisions 
regarding non-computable sets  can be made. The general properties of $\gD_2$ 
sets can be found in \cite{shoenfield}.

\bthm
\label{delta2}
Let $A \subseteq \oN$ be a $\gD_2$ set of natural numbers. Assume $x_1,x_2,...$
is a sequence of i.i.d real random variables with mean $\gm$ and variance 
$\gs^2<\infty$. Then there exist a
recursive decision procedure $F:R^{<\go} \rarrow \{0,1\}$ such that almost surely (with respect to realizations of the sequence) 
\beq
\lim_{n \rarrow \infty} F\left(\left<x_1,...,x_n \right> \right)=1 
\iff \gm \in A
\eeq
\ethm
Note that unlike the Cover Koplowitz theorem we do not require that the set 
$A$ admits a recursive enumeration. This strengthens the effective version of 
the Cover-Koplowitz theorem to $\gD_2$ sets of integers.

Proof: Our decision procedure consists of two steps. We First decide whether 
$\gm$ is an integer based on an approach similar to the Cover-Koplowitz 
theorem using that the set $\oN$ admits a recursive numeration. 
Then we rely on the law of iterated logarithms to obtain an 
estimate of $\gm$ and exploit the definition of $A$ as a $\gD_2$ set.
To test whether $\gm \in \oN$ let 
$
\barx_N= \frac{1}{N} \sum_{n=1}^N x_n \\
$
be the sample mean and let 
%\beq
$\bargs_N=\frac{1}{N} \sum_{n=1}^N (x_n-\barx_N)^2
$
%\eeq
be the sample variance.
Almost surely $\barx_N \rarrow \gm$ and 
$\bargs \rarrow \gs^2$.
Furthermore by the law of iterated logarithms almost surely only finitely many
times
%\beq
$
|\gm-\barx_N|>\sqrt{\frac{\gs^2 \log \log N}{N}}
$
%\eeq
and for every $\gre>0$ almost surely only finitely many times 
\beq
\gs^2>(1+\gre)\bargs_N
\eeq
Fix $\gre>0$.
Let $\gd_N=\sqrt{\frac{(1+\gre)\bargs_N \log \log N}{N}}$
and
%\beq
$
\gmh_N=\round(\barx_N).
$
%\eeq  
If $\gm \in \oN$ then with probability one only finitely many times
$\gmh_N \neq \gm$, and except finitely many times this holds whenever 
%\[
$
|\gmh_N-\barx_N|<\gd_N
$
%\]
If on the other hand $\gm \nin \oN$ then almost surely except finitely many 
$N$'s $|\gm-\gmh_N|>\gd_N$ and therefore almost surely except finitely many 
times
%\[
$
|\gmh_N-\barx_N|>\gd_N
$
%\]
Let $\left<d_n:n=0,1,...\right>$ be a sequence of decisions for the hypothesis
$H^{\oN}_0: \gm \in {\oN}$ given by 
\beq
d_n=\left\{
\bea{ll}
1 & \hbox{if} \ \ |\gmh_n-\barx_n|\le \gd_n \\
0 & \hbox{if} \ \ |\gmh_n-\barx_n|>\gd_n
\ena
\right.
\eeq
By the discussion above almost surely $d_n$ is correct except finitely many 
times. 
If $d_n =0$ decide $\gm \nin A$. Otherwise we assume that $\gm \in \oN$ and 
 almost surely $\gmh_n=\gm$ for all 
but finitely many $n$'s. Hence if $d_n=1$ assume that $\gm_n \in \oN$. 
We now define a procedure for testing whether $\gmh_n \in A$. 
To that end recall that since $A$
is $\gD_2$ there are recursive relations $\phi(m,k,n), \psi(m,k,n)$ such that
\beq
\label{Aformulas}
\bea{lcl}
n \in A &\iff& \exists m \forall k [\phi(m,k,n)] \\
n \nin A &\iff& \exists m \forall k [\psi(m,k,n)] 
\ena
\eeq
\bdf
An $m_0$ such that $\forall k \phi(m_0,k,n)$ is called a witness for 
$n \in A$.
An $m_0$ such that $\forall k \psi(m_0,k,n)$ is called a witness for 
$n \nin A$.
$m_0$ decides $n \in A$ if $m_0$ is a witness for $n \in A$ or for 
$n \nin A$.
\edf
Obviously by (\ref{Aformulas}) for each $n$ either there is a witness 
that $n \in A$ or there is a witness for $n \nin A$. 
Therefore given $n$ and $m$ if we compute 
$\left< \left(\phi(m,k,n),\psi(m,k,n)\right), k=1,2,... \right>$ 
we must obtain some $k$ where either $\phi(m,k,n)$ fails or $\psi(m,k,n)$
fails. Moreover by definition there is a least $m$ where one of the formulas 
holds for all $k$. 

Let $\gmh_N$ be the estimate of $\gm$ given that $d_N=1$.
Let $m_0$ be the minimal $m$ such that only one of the formulas fails for 
some $k$ (assume without loss of generality this is $\phi$). Then with probability one except 
finitely many  $N$'s
\beq
\forall k<N [\phi(m_0,k,\gmh_N)]
\eeq
and for sufficiently large $N$ $m_0$ is the minimal one.
This suggests the following decision procedure:
Assume that we have decided that at stage $N$ that $\gm \in \oN$ ($d_N=1$).
Decide $e_N=1$ if the minimal $m$ such that 
\beq
\forall k<N [\phi(m,k,\gmh_N)]
\eeq
is smaller than the minimal $m$ such that 
\beq
\forall k<N [\psi(m,k,\gmh_N)].
\eeq
By the discussion above, if $\gm \in A$ then $e_N=1$  
except finitely many $N$'s and if $\gm \nin A$, $e_N=0$ 
except finitely many $N$'s. This ends the proof.

Finally we comment that if we replace $\oN$ by any countable recursive sequence
of recursive reals $S=\left<s_n: n \in \oN \right>$
and $A \subseteq \oN$ is $\gD_2$ then we can extend the above result to test 
whether $\gm \in \left<s_n: n \in A \right>$. However similarly to Cover's 
result if the closure of $S$ is uncountable then there is a measure $0$ 
subset of $\oR \backslash S$ on which the test might fail. 

\bibliographystyle{plainnat}
%\bibliography{cover}

\end{document}